\documentclass[lettersize,journal]{IEEEtran}
\usepackage{amsmath,amsfonts}
\usepackage{algorithmic}
\usepackage{array}
\usepackage[caption=false,font=normalsize,labelfont=sf,textfont=sf]{subfig}
\usepackage{textcomp}
\usepackage{stfloats}
\usepackage{url}
\usepackage{verbatim}
\usepackage{graphicx}
\def\BibTeX{{\rm B\kern-.05em{\sc i\kern-.025em b}\kern-.08em
    T\kern-.1667em\lower.7ex\hbox{E}\kern-.125emX}}
\usepackage{balance}

\usepackage{booktabs}
\usepackage{multirow}
\usepackage{svg}
\usepackage{amsmath}
\usepackage{enumitem}
\setlist[itemize]{nosep}
\usepackage{float}
\usepackage{bm}
\usepackage{url} 
\usepackage[utf8]{inputenc}
\usepackage{graphicx}
\usepackage{xspace}
\usepackage{xcolor}
\usepackage{enumitem}
\usepackage{array}

\begin{document}

\title{Energy Consumption in Next Generation Radio Access Networks}

\author{\IEEEauthorblockN{Urooj Tariq, Rishu Raj, Merim Dzaferagic and Daniel Kilper} \\CONNECT Centre, Trinity College Dublin, Ireland.\\
\small Email: \{tariqu, rajr, merim.dzaferagic, dan.kilper\}@tcd.ie
\vspace{-6mm}}

\maketitle

\begin{abstract} 
Historically, in macro-cell-based networks, the radio access network (RAN) portion accounts for the largest share of energy consumption in mobile networks. With greater densification, the transition to small cells and increased virtualization of the radio processing functions, the relative energy use of the different network segments is changing. At the same time, open radio access networks (O-RAN) have emerged as a promising approach to support this evolution through open interfaces that enable a multi-vendor environment, support for hierarchical intelligent controls, and simplified, cost-effective radio units that facilitate large-scale deployments. This paper examines the impact of these changes on energy consumption in next-generation RAN architectures through transaction-based energy models. Transaction-based network energy models were developed in 2009 to understand the evolution of energy use in networks. Here we update these models with today's hardware and apply recently developed models for emerging architectures, considering current access point densification trends. Results indicate that efficiency improvements have followed predicted trends, which were largely based on semiconductor Moore's Law-related evolution. Emerging RAN architectures show improved energy per bit, although the processing energy becomes the largest contributor. Densification increases total energy consumption, but the relative contributions of processing and transport energy depend strongly on the location of the baseband processing. 
\end{abstract}

\begin{IEEEkeywords}
Energy Consumption, Radio Access Networks, O-RAN, Energy Modeling.
\end{IEEEkeywords}

\section{Introduction}
\IEEEPARstart{C}{ommunication}\label{Section I}
networks have evolved from systems primarily designed for human communication to infrastructures increasingly dominated by connected devices. While billions of people continue to use the network for everyday services, an even larger number of devices are now connected, generating and exchanging data autonomously. This evolution has driven a transition to a world in which the number of connected devices exceeds the human population, creating significant challenges in terms of scalability, latency, and energy efficiency. According to estimates, between 25 and 50 billion devices are being used, producing 79.4 zettabytes of data annually \cite{qi2020integration}. In 2023, 65.7\% of the global population \cite{badshah2024optimizing} used the Internet for applications with diverse requirements, including high data rates, low latency, and high mobility.

The exponential growth of Internet traffic has frequently drawn concern over the potential for a rapid expansion of equipment deployments and corresponding increases in energy use. Much attention was focused on this more than a decade ago, when it was first recognized that the energy efficiency improvements in network equipment were slowing and not keeping up with traffic growth \cite{9428552}. In order for traffic to scale exponentially, the network equipment efficiency, measured in energy per bit (transported), must reduce exponentially. Historically, this was achieved by increasing transmission capacity with little increase in power. These gains slowed as wireless and, subsequently, wireline systems approached Shannon capacity limits, while processor scaling was constrained by the 'one volt wall' and other limitations on the efficiency improvements associated with Moore's Law \cite{10602753}. Earlier studies therefore predicted increasing network equipment energy use. In fact, the most recent generation of 800G optical transceivers is being offered with liquid cooling in order to address the high thermal densities. The impact of these equipment efficiency trends on actual commercial network energy use is complicated due to the relative mix of legacy and new equipment deployments and efficiencies in business operations specific to each operator. Nevertheless, a number of studies have found evidence of industry-wide network sector increases in electricity use, even growing faster than computers and data centers (DCs) \cite{11404237}. 

\begin{figure}[!t]
    \centering
    \includegraphics[width=6.5cm]{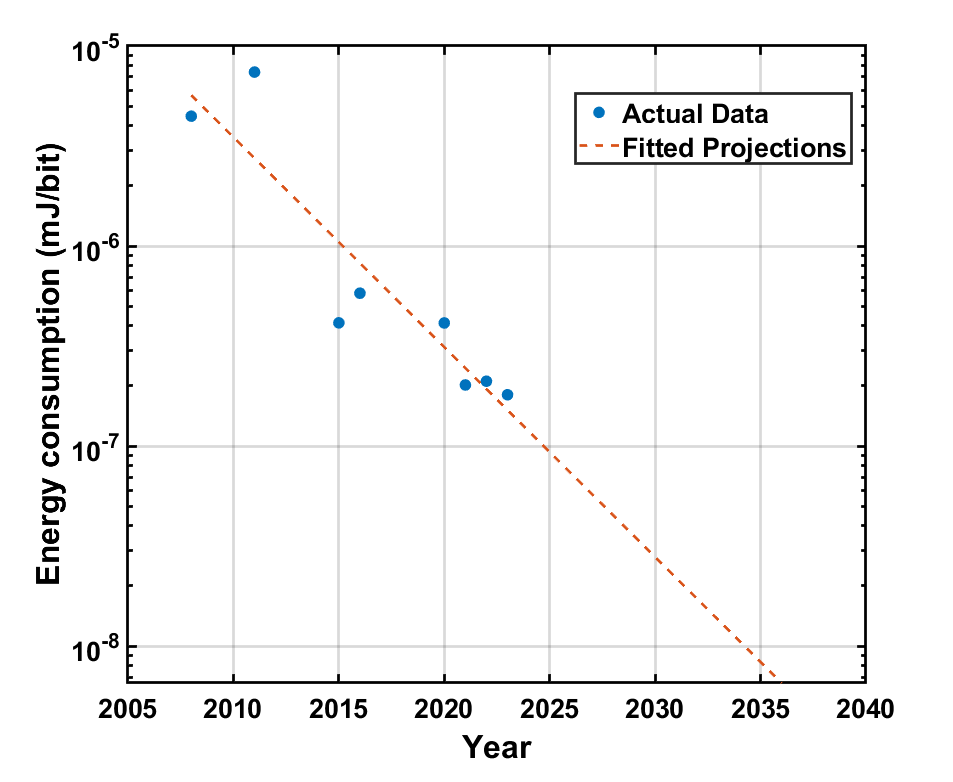}
    \vspace{-4mm}
      \caption{Energy consumption per user in access switches over time with a technology improvement rate of 20\%.}     
    \label{fig:cisco trend}
    \vspace{-4mm}
\end{figure}

The global climate crisis likewise has significantly shifted our attention towards scrutinizing, discussing, and reducing the energy consumption and carbon footprint of the information and communication technology (ICT) sector, which was estimated at 1.27 Gt of carbon dioxide emissions ($\text{CO}_\text{2e}$), or 2.3 \% of global emissions in 2020. These trends motivate re-examining whether previous predictions of network energy-use evolution have borne out. Assuming semiconductor scaling is the dominant contributor to energy use, an exponential time-dependent model has been used for the prediction of energy consumption (or energy per bit) expressed as $E(t)  = E_\text{0}(1-\mu)^ {t-t_0}$ where $\mu$  is the annual rate of technological improvement, $E_{0}$ is the energy consumption of the access switches at $t_{0}$. Previous studies predicted $\mu=0.2$ per year \cite{7247606}. 
Ethernet switches are an important category of network equipment that depend strongly on semiconductor scaling and have not seen a significant shift in their architecture or use. We therefore analyse a family of Ethernet switches from a common vendor to limit vendor- and design-dependent variation. In Fig. \ref{fig:cisco trend}, we compare the exponential trend with the energy consumption of access and aggregation switches based on public datasheets from the year 2008 to 2023. We also show that the log-domain fit gives $R^2=0.848$, corroborating the previous prediction and supporting $\mu=0.2$ as a representative value.

Understanding the energy consumption of network equipment is complex and has been approached through different methods. The most commonly used method is a so-called `top-down' economic-based model that estimates the overall network power consumption from publicly reported network operator business operations figures or equipment sales figures \cite{10746340}. This gives an accurate indication of energy use for the sector overall, but does not provide insights into how the network is configured or operated, which might include legacy equipment or redundant deployments. To better understand network efficiency and energy use, bottom-up methods can be used, which model architectures based on available equipment or equipment models \cite{11162430}. These methods can be used to study a wide range of network optimization and techno-economic analyses. One such bottom-up approach is the transaction-based modeling technique, which models data traffic flow through a network based on mean quantities such as the mean hop counts and typical equipment configurations. Similar to lifecycle analysis methods for goods shipping and transportation, transaction-based network models follow the data path of a service and estimate the accumulated energy contribution of each path segment. This type of modeling has been applied to individual services, variations in the traffic volume of the network \cite{kilper2010power}, power consumption in optical IP networks 
and wired and wireless access networks \cite{baliga2011energy}.

Using transaction-based methods, wireless access networks were shown to be the most energy-consuming segment of the overall global network infrastructure. In order to continue to increase capacity, wireless networks must rely on using more bandwidth (spectrum) and spatial diversity by reducing the cell size or distance to the access point. Both approaches require more energy. The introduction of centralized RAN (C-RAN) architectures is in part done to reduce the energy requirements of access point densification by moving the baseband processing (BBP) to more efficient, shared processing locations in DCs. It also allows for greater use of virtualized network functions. Numerous studies of C-RAN architectures have considered the optimization of both wireless and compute resources \cite{11124199}. With commercial equipment becoming available for open-RAN (O-RAN) deployments, energy optimization of the wireless network through C-RAN and virtualization has become a key use case and stimulated renewed attention. It also motivates the study of such architectures using transaction-based methods and revisiting previous studies with updated equipment and new architectures. Recently \cite{11162430}, this approach was used to study the power consumption in O-RAN architectures for different network topologies and nodal fanouts.

Using these new transaction-based models for 5G architectures, including the full front- and back-haul networks, it is possible to analyze their evolution and connection to previous architectures and past predictions of energy use.  Here we perform such an analysis by first considering both the processing energy at different nodal units and the transmission energy associated with high-capacity evolved common public radio interface (eCPRI) data \cite{11124199}. The eCPRI protocol used in 5G carries digitized radio signals along with control, management, and synchronization information. We update previous models \cite{baliga2011energy} by using the most recent generation of network equipment, and then perform a comparative evaluation between conventional access network designs and the new O-RAN-based architectures for different user access rates.  From an access technology perspective, the energy benefits of the new centralized architectures are readily apparent, showing efficiency similar to passive optical networks (PONs) at access rates of 100 Mbps. A full understanding of these new architectures, however, is complicated by the distribution of processing equipment throughout the network. Therefore, we further examine the effect of placing BBP at different centralization points in the RAN architecture and evaluate performance under varying levels of network densification for a fixed number of users, providing a framework for understanding the impact of densification and its trade-offs with respect to energy use, which is the critical factor for energy use in these networks. The results highlight how the placement of BBP and the degree of network densification influence overall energy efficiency. The results show that in these new small cell architectures, the processing energy, rather than the radio unit energy, dominates total consumption. The transaction-based models capture the corresponding architectural trade-offs and show that centralized architectures exhibit the potential to achieve significant savings when resources are efficiently shared.

\section{RAN Architectures}

Radio access networks (RANs) have evolved significantly across successive generations of wireless communication systems to meet growing connectivity demands. Early implementations were characterized by closed, hardware-dependent architectures and used macro-cell base stations with co-located RF and radio processing equipment, often referred to as distributed RAN (D-RAN) today. This architecture was efficient for large macrocell deployments aggregating radio traffic over 10-20 km distances and serving an average of 1000 users. Later, 4G LTE benefited from a transition to remote radio units, for which the radio equipment was connected to the baseband unit (BBU) through an optical fiber connection (from the top of the cell tower to the base), saving significant RF transmission power. Despite LTE’s energy gains, architectural constraints at the cell site remained the bottleneck for scaling and cost due to D-RAN limitations, as each eNodeB operates independently and relies on proprietary systems.

To overcome these issues, the C-RAN architecture was introduced, which decoupled radio units (RUs) from BBUs using the long reach of the fiber connection to distribute RUs at remote cell sites while pooling BBUs in a centralised location. This arrangement is designed to allow multiple RUs to share processing resources, improving efficiency and reducing operational costs. Centralised BBUs also enable more flexible scaling, since additional RUs can be deployed without replicating processing units at each site. An optical fronthaul network, typically based on the CPRI protocol over fibre, connects the distributed RUs with the centralised BBU pool. Virtualised RAN (vRAN) and Cloud-RAN ease vendor lock-in and let the same hardware be shared across RAN variants, so deployments can be more flexible, easier to scale, and quicker to adapt \cite{11124199}. Operationally, vRAN emerged in late 4G/LTE deployments and has become more prevalent with 5G, where wider bandwidths and larger-scale MIMO make pooling and automation particularly valuable. Cloud-RAN is largely a 5G-era development, as operators relocate functions of distributed units (DUs) and centralised units (CUs) to edge DCs or centralized clouds to improve utilization and orchestration \cite{11517492}. vRAN virtualizes DU/CU processing on commercial off-the-shelf (COTS) platforms to improve utilization and agility, but it does not guarantee multi-vendor openness or higher efficiency because CPUs/GPUs may not outperform dedicated radio-processing ASICs. 

\begin{figure*}[!t]
    \centering
    \includegraphics[width=6.1in]{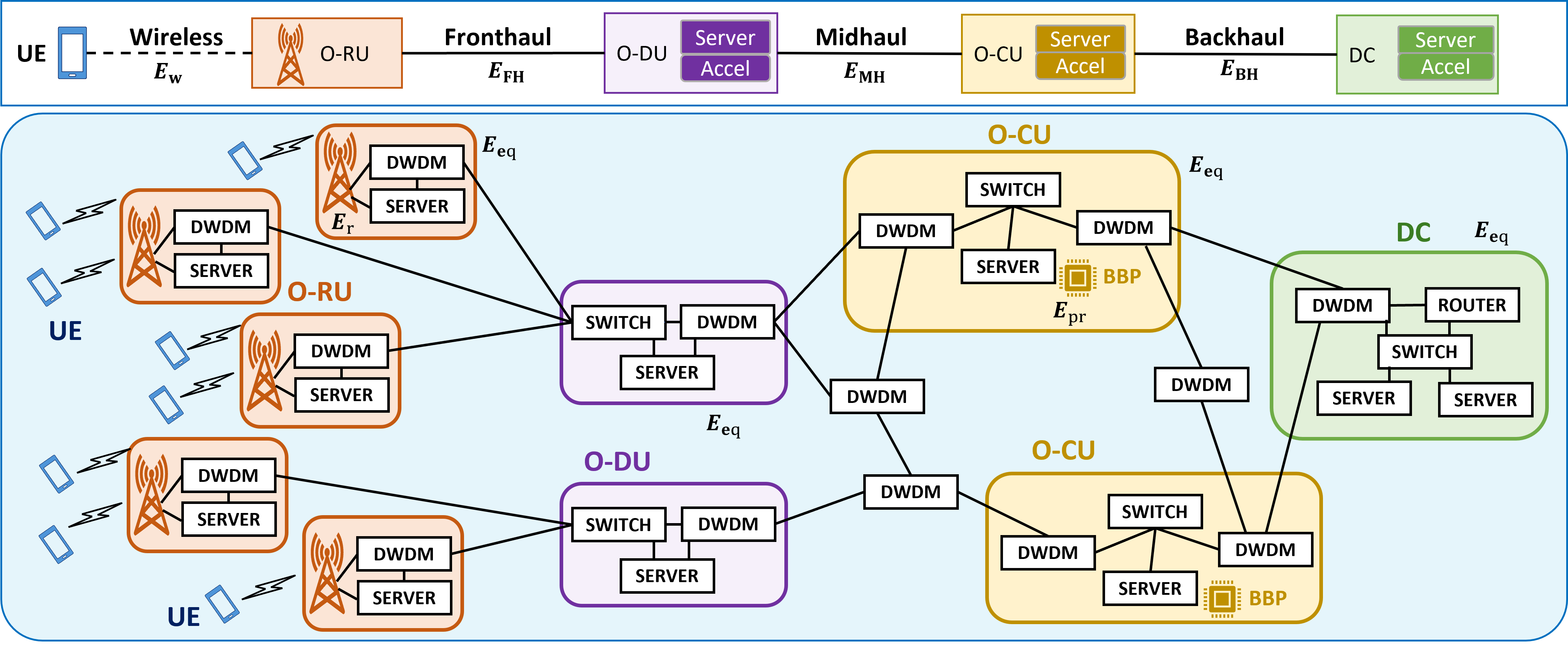}
    \vspace{-4mm}
    \caption{O-RAN architecture showing baseband processing (BBP) at O-CU.} 
    \label{fig:block_diagram}
    \vspace{-4mm}
\end{figure*}

Building on virtualization, O-RAN was developed as a fully open and disaggregated framework for 5G RAN. Unlike standardised architectures focused on functionality, O-RAN disaggregates hardware and software, enabling network functions to run on COTS hardware. This reduces dependence on proprietary systems while enhancing flexibility. The O-RAN architecture is composed of open radio units (O-RUs), open distributed units (O-DUs), and open central units (O-CUs) (Fig. \ref{fig:block_diagram}). O-RAN makes energy a controllable resource \cite{10592054} by disaggregating O-RU/O-DU/O-CU and exposing policy controls through near-/non-real time (RT) RAN intelligent controllers (RICs) for cell muting, carrier thinning, MIMO layer capping, and deep sleep orchestration. The RIC uses modular applications known as xApps and rApps, which are vendor-developed software modules that operate within the RIC platform and perform specific control tasks \cite{11124199}. Building on this programmability, energy management has emerged as a key use case for O-RAN, with several studies investigating the use of xApps and rApps to develop RIC-based solutions for energy-aware RAN operation under network-performance constraints \cite{11124199, 10835138}.

Mobile networks using a C-RAN architecture can be organized into fronthaul, midhaul, or backhaul networks that extend across the edge, metro, regional, and longhaul hierarchical areas. Fronthaul connections are typically point-to-point links between the O-RUs and the nearest O-DU/O-CU aggregation and processing nodes. Midhaul carries partially processed signals, typically between O-DU and O-CU, while backhaul carries processed user traffic deeper into the network. Together, these segments form the x-haul network, with traffic increasingly multiplexed over dense wavelength-division multiplexing (DWDM) and routed through core gateways as it is aggregated.
To allow for the flexibility and virtualization of radio processing at different locations in the network, 3GPP introduced functional splits, which divide the radio processing into its various steps separated by standardized split points. Processing to a given split point consumes processing power but reduces transport capacity by converting high-rate digitized radio traffic toward baseband user data. This creates a fundamental processing–transport tradeoff \cite{11124199}. 3GPP specifies eight split options for the distribution of the 5G gnodeB (gNB) across the network \cite{11296907}. In O-RAN, Split 7.2 is recommended between the RU and DU, which transports the digitized RF signal in blocks. Many combinations of processing splits can be used among the different x-haul network connections, allowing for flexible trade-offs between processing, transport, and functionality that need to be captured in order to understand the consequences on network energy use.  

\section{Energy Use Models}\label{Section III}

\subsection{Access Networks}
Various power consumption models exist in the literature for different network architectures as discussed in Section \ref{Section I}. In order to explore the evolution of network energy use over the past decade, we start with the transaction-based models introduced in 2011\cite{baliga2011energy}, where the authors analyze both wired and wireless access network technologies. The energy consumption per user is given as \cite{baliga2011energy},
 \begin{equation}
\begin{aligned}
   E_{\text{U}} = \frac{2}{R_{\text{U}}}\left(\frac{P_\text{TU}}{N_\text{TU}}+\frac{P_\text{RN}}{N_\text{RN}}+\frac{P_\text{CPE}}{2}\right)
\end{aligned}
\label{eq:1}
\end{equation}
where $R_{\text{U}}$ is the user access rate, $N_\text{TU}$ and $N_\text{RN}$ are the average number of users served by the terminal unit and remote node, respectively, and $P_\text{CPE}$, $P_\text{RN}$ and $P_\text{TU}$ are the power of the customer premises equipment (CPE), the remote node and the terminal unit in the central office, respectively.  In the original work, this model was generalized to include PONs, point-to-point (PtP) networks, Fiber to the node (FTTN), and worldwide 
interoperability for microwave access (WiMAX).

\begin{figure}[!t]
    \centering
    \includegraphics[width=8cm]{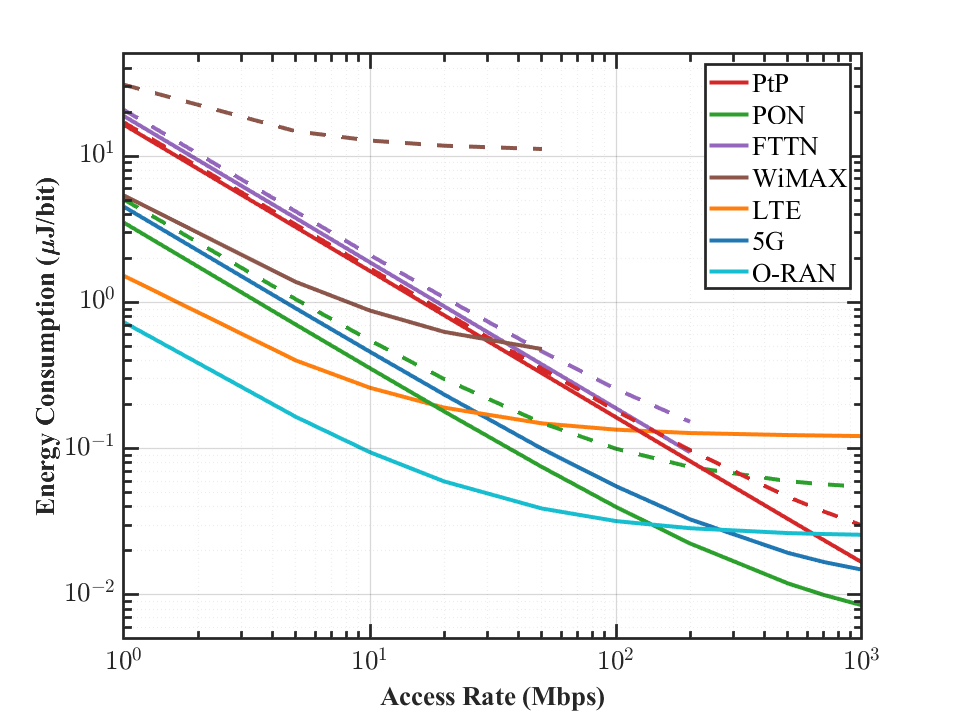}
    \vspace{-4mm}
    \caption{Energy consumption of various access network technologies. The curves for PON, FTTN, PtP, and WiMAX are generated using the equipment in \cite{baliga2011energy} (dashed lines) and also the latest network equipment (solid lines). LTE, 5G and O-RAN are not studied in \cite{baliga2011energy}.}
    \label{fig:power_comparison_tucker_2}
    \vspace{-2mm}
\end{figure}

In Fig.~\ref{fig:power_comparison_tucker_2}, we reproduce the prior energy consumption analysis of PON, FTTN, PtP, and WiMAX~\cite{baliga2011energy}. The legacy equipment parameters are adopted from~\cite{baliga2011energy}, while more recent commercially available hardware is used to update the equipment assumptions. We further extend the comparison to LTE, 5G, and O-RAN, which were not considered. For the updated equipment, PtP uses a Perle S-10GRT-SFP media converter as the CPE and no active remote node, while PON uses a UFiber loco optucal network unit (ONU) as the CPE and a ZTE ZXA10 C320 optical line terminal (OLT) as the terminal unit. FTTN uses a Zyxel GM4100-B0 VDSL2 modem as the CPE and a Huawei MA5818-24P DSLAM at the remote node, whereas WiMAX uses an Alvarion BreezeMAX USB 200/Zyxel MAX-200M1 as the CPE and a Motorola WAP 450 Series base station at the remote node. For LTE, the CPE is a ZTE MF823 modem, and the radio-access power follows the EARTH 2012 base-station model, while for 5G, a mobile device is used as the CPE and a  Benetel radio~\cite{11162430} represents the radio node. A Cisco Catalyst 1300 Ethernet switch is used as the terminal unit for PtP, FTTN, WiMAX, LTE, and 5G. The corresponding CPE, radio, transport, and processing equipment used for the O-RAN configurations are detailed in Table~\ref{tab1}.

Across the evaluated technologies, the high energy consumption per bit at low access rates is mainly caused by poor utilisation of equipment whose power consumption is largely fixed or only weakly dependent on traffic. When only a small amount of traffic is carried, this baseline power is attributed to relatively few transmitted bits, resulting in a high energy cost per bit. As the access rate increases, the same fixed contribution is shared over a larger traffic volume, causing the energy per bit to decrease rapidly. This behaviour is consistent with previous studies, which show that fixed equipment power dominates at low access rates and that underutilisation substantially reduces energy efficiency~\cite{baliga2011energy}. In the 5G configuration, the comparatively large fixed radio and BBP contribution gives a higher energy per bit than LTE at low access rates. However, its higher supported capacity allows this fixed contribution to be shared over a larger traffic volume as the access rate increases.

The differences in the energy consumption of these access technologies become more evident at medium access rates, where resource sharing, network architecture, and equipment characteristics increasingly determine the relative performance. WiMAX maintains comparatively high energy-per-bit values because of the relatively high power consumption and limited shared capacity of its base station infrastructure. As the access rate per user increases, fewer users can share the capacity of a given base station sector, so a larger fraction of the radio infrastructure power is attributed to each user and additional sectors or base stations may eventually be required to support the same user population. FTTN requires very-high-speed digital subscriber line (VDSL) switching equipment at the remote node together with a VDSL2 modem at the customer premises, and therefore retains an additional powered network element that increases its energy consumption relative to architectures without an active remote node. In contrast, PON and PtP become increasingly energy efficient as the access rate increases, consistent with their favorable high rate behaviour~\cite{baliga2011energy}. For PON, the passive splitter introduces no additional power consumption in the distribution network while the OLT capacity is shared among multiple users, allowing the power of the active equipment to be distributed efficiently as traffic increases. PtP, in turn, provides a dedicated high-capacity optical connection to each user without an active remote aggregation node, avoiding the additional power required by remote nodes in FTTN. These architectural differences explain why the optical technologies become increasingly favorable as higher access rates are required.

LTE performs relatively well at lower rates, but its base station contains both largely fixed radio/baseband components and a load-dependent power amplifier contribution. As the fixed component is increasingly distributed, the remaining load-dependent contribution causes the energy per bit to approach a nearly constant level. The 5G configuration continues to decrease over a wider range of access rates because its higher supported capacity allows its comparatively larger fixed radio and BBP contribution to be distributed over a greater traffic volume, making it increasingly favourable relative to LTE at higher rates. In deployment terms, these trends show that the preferred access technology depends not only on nominal equipment power, but also on how effectively infrastructure can be shared and how much additional equipment must be provisioned as the required per-user rate increases.

The crossover at higher access rates reflects the balance between fixed equipment power, available capacity, and shared resources. As the carried traffic increases, fixed equipment power is distributed over a larger traffic volume, reducing the energy consumed per bit. This effect is particularly visible for PtP, FTTN, and PON. For PON, the energy per bit is initially higher than that of LTE and O-RAN, but decreases rapidly as the shared OLT capacity is better utilized, and the passive distribution network introduces no additional power contribution at the remote nodes. Likewise, the 5G configuration becomes more efficient as the access rate increases because its comparatively high fixed radio and BBP power is distributed over a larger traffic volume. LTE, in contrast, approaches an energy-per-bit floor as the fixed base-station power becomes less significant at high traffic rates, while transmitting and processing each additional bit still consumes energy. Therefore, the energy per bit gradually levels off instead of continuing to decrease. O-RAN achieves comparatively low energy per bit due to shared processing resources, and high-capacity radio and transport components. At higher access rates, the fixed power contribution becomes small when distributed over a large traffic volume, while each additional bit still requires radio transmission, processing, and transport. These remaining traffic-dependent contributions result in lower energy consumption that changes only gradually when the access rate is high, causing the O-RAN energy consumption to approach an approximate floor. Overall, technologies with substantial fixed power relative to the carried traffic can appear inefficient under light load but become increasingly favorable as utilization improves.

In essence, the results can be viewed across three operating regions. At low access rates, static and idle equipment power has the strongest influence. At medium rates, sharing and architectural differences determine how efficiently the network supports increasing traffic. At high access rates, available capacity and load-dependent equipment power increasingly determine the energy cost per bit. Consequently, the relative ranking of the technologies changes with access rate rather than one technology remaining uniformly preferable, which is also observed in previous studies~\cite{baliga2011energy}.
Note that this analysis only considers the access portion of the network \cite{baliga2011energy} and does not incorporate the energy consumed in the transmission of data over the larger network connecting to DCs. Furthermore, the additional energy used for thermal management, protection resources, or mobility coverage is not included here \cite{baliga2011energy}. Thus, this analysis provides a basic, coarse comparison of the energy use of different access technologies without accounting for the complexities of their deployment requirements and operation.

\begin{figure}[!t]
    \centering
    \includegraphics[width=6.7cm]{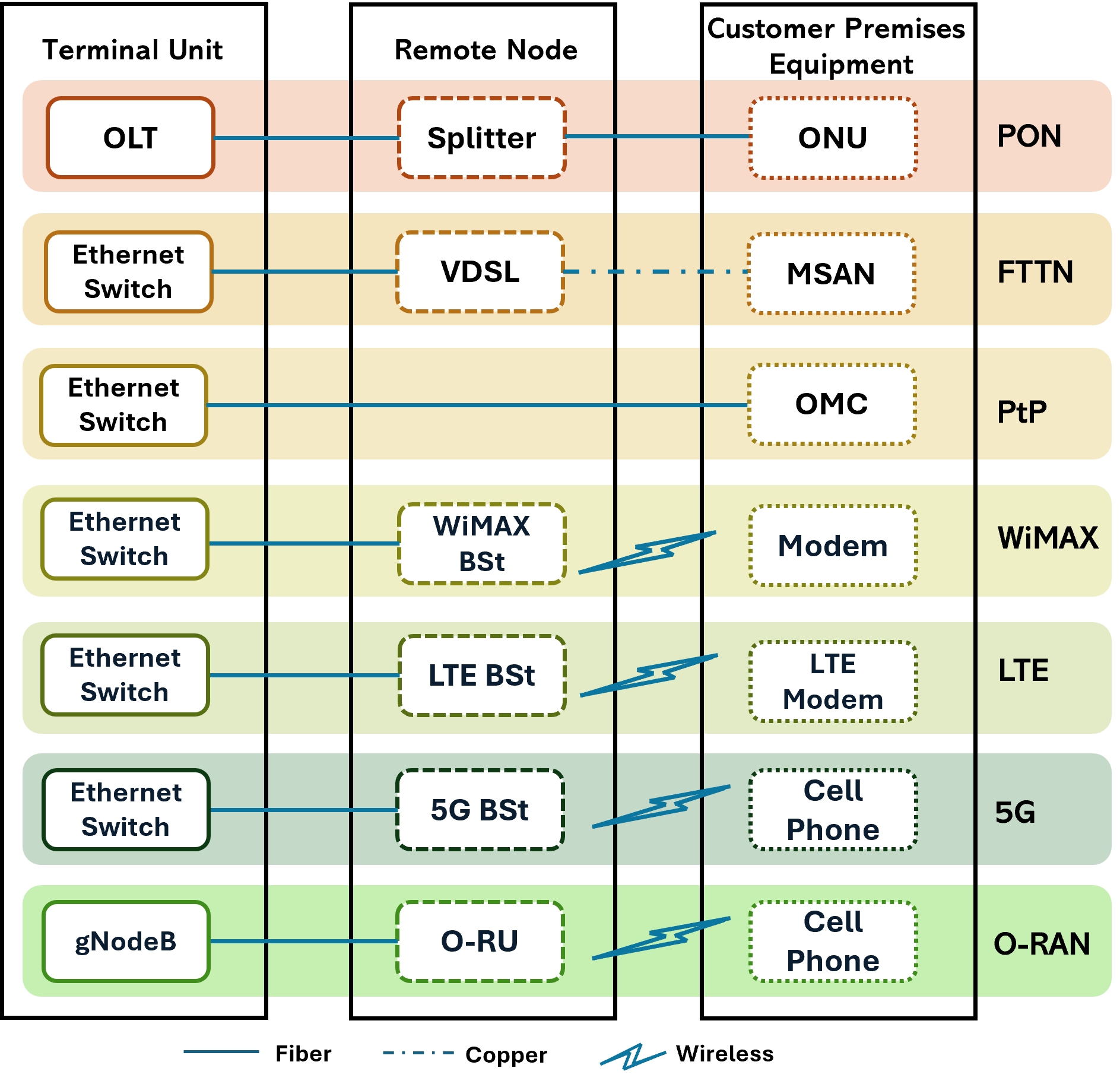}
    \vspace{-4mm}
    \caption{Access network architecture for various technological platforms.}
        \label{fig:network_schematics}
        \vspace{-4mm}
\end{figure}

\subsection{Energy Consumption Model}

Given that recent radio architectures rely on network-scale efficiency enhancements, the coarse model shown in Fig. \ref{fig:network_schematics} needs to be enhanced to include the overall network between the access point and the DC.  We apply previously developed network-scale transaction-based models to recent 5G C-RAN uplink scenarios, including O-RAN.
Here, an `over-provisioning' factor is used to account for provisioning above that needed to handle the mean traffic, for example, to be able to accommodate diurnal and other variable traffic patterns or to support survivability functions \cite{11162430}. An  `overhead' modifier is used to account for cooling, power distribution losses, and other network equipment inefficiencies \cite{baliga2011energy}. A coverage factor is used to account for radio network deployments needed to cover a wide area with a desired density with respect to the average number of users. Today's RAN extends from the RU to the DU and CU located in nearby DCs or processing sites, and application data might continue to a regional DC where the user application processing takes place. We look at the end-to-end architecture and identify the different components that contribute to the overall energy consumption in the network. For a centralized RAN, this network includes fronthaul, midhaul, and backhaul transmission links. Moreover, at each nodal unit in the network, energy is consumed if processing operations are performed on the user traffic. Therefore, the total energy consumption per user bit is expressed as
\vspace{-2mm}

\begin{equation}
\vspace{-1mm}
\begin{aligned}
E_{\mathrm{T}}
={}&
\underbrace{E_{\mathrm{ra}}}_{\text{radio}}
+
\underbrace{
\sum_{u\in\mathcal U}
\Phi_u\frac{P_{\mathrm{C},u}}{C_{\mathrm{C},u}}
}_{\text{processing}}
+
\underbrace{
\sum_{u\in\mathcal U}
\Psi_u\frac{P_{\mathrm{I},u}}{C_{\mathrm{I},u}}
+
\sum_{\ell\in\mathcal L}
\Omega_\ell\frac{P_\ell}{C_\ell}
}_{\text{transport}}.
\end{aligned}
\label{eq:total_energy}
\end{equation}
Here $E_{\mathrm{ra}}$ denotes the radio energy consumption and is given by $E_{\mathrm{ra}}=E_w+N_rP_r/C_{\mathrm{U}}$, where $E_w$ is the energy consumed in the wireless transfer of a user bit to the serving RU, $N_r$ is the number of RUs, $P_r$ is the power consumption of the radio equipment at each RU, and $C_{\mathrm{U}}$ is the aggregate user traffic rate. The second term represents the processing energy ($E_{\mathrm{pr}}$) consumed for BBP. The set $\mathcal{U}$ contains the RAN nodes at which processing may be performed, including the RU, DU, CU, and regional DC. For node $u$, $P_{\mathrm{C},u}$ denotes the aggregate power consumption of the active processor cores used for BBP, while $C_{\mathrm{C},u}$ denotes the processing capacity of the server. The coefficient $\Phi_u$ combines the dimensionless factors associated with processing at node $u$, including server over-provisioning, processing overhead, coverage-related scaling, and the ratio of the network side traffic rate to the aggregate user traffic rate. For a node at which BBP is not performed, $\Phi_u=0$~\cite{11162430}.

The third and fourth terms together constitute the transport energy ($E_{\mathrm{tr}}$). The third term accounts for the energy consumed by the equipment associated with the network nodes, whereas the fourth term accounts for the energy consumption of the equipment associated with the x-haul segments traversed by the traffic. For the third term, $P_{\mathrm{I},u}$ and $C_{\mathrm{I},u}$ denote the power consumption and traffic handling capacity, respectively, of the network interface or line-system equipment at node $u$. The coefficient $\Psi_u$ combines the corresponding over-provisioning, equipment overhead, and traffic scaling factors. Its value depends on the location of node $u$ relative to the BBP point. Before BBP, the transport equipment carries the higher-rate radio-domain traffic, whereas after BBP it carries the corresponding user-domain packet traffic. For the fourth term, the set $\mathcal{L}$ contains the x-haul segments traversed by the traffic, including the applicable fronthaul, midhaul, and backhaul segments. For segment $\ell$, $P_\ell$ and $C_\ell$ denote the corresponding transmission equipment power and capacity parameters. The coefficient $\Omega_\ell$ includes contributions from different switches, DWDM links, and routers for the x-haul segments. It incorporates the segment-dependent over-provisioning, equipment overhead, traffic scaling, and the numbers and types of transmission stages traversed within that segment. Moreover, $\Omega_\ell$ also captures the dependence of the x-haul energy on whether the segment carries traffic before or after BBP~\cite{11162430}. Hence, $\Phi_u$, $\Psi_u$, and $\Omega_\ell$ represent the aggregate scaling factors for BBP functions, nodal transport, and x-haul transport, respectively, in \eqref{eq:total_energy}.

\section{RAN Scenarios}
Details of the network equipment used in this analysis are given in Table \ref{tab2}. We assume that the average monthly data consumption of a user is 10 Gigabytes. Typical of RAN deployments, the servers at distributed radio locations in the network have 4 cores with a total BBP capacity of 1 Gbps each, whereas the servers in a DC have 20 cores, which gives a cumulative BBP capacity of 5 Gbps of fronthaul data per server. The thermal design power per core for the servers at RU, DU and CU is 6 W, and for those at the DC, it is 5.5 W. The energy expended in the user equipment (UE) is \(E_{\mathrm{w}}\) = 25 nJ/bit. These design parameters are fixed unless mentioned otherwise~\cite{11162430}.

\begin{table}[t]
  \centering
  \refstepcounter{table}
  \label{tab2}
  {\footnotesize
\textsc{TABLE~\thetable.}~
\textsc{Network Equipment Specifications \cite{11162430}}
\par}
  \vspace{3pt}
  \resizebox{\columnwidth}{!}{%
  \begin{tabular}{l l c c}
    \hline
    Device & Model & Rated Power (W) & Capacity (Gbps) \\
    \hline
    Radio       & Benetel 650          & 110  & 11  \\
   Access Switch& Cisco Catalyst 1300  & 86.7 & 480 \\
    Router     & Cisco 8000           & 172  & 3200   \\
    DWDM        & 1FINITY T600         & 4265 & 9600  \\
     Core Switch & Cisco 9600           & 3000 & 25600    \\
    \hline
  \end{tabular}%
  } \label{tab1}
  \vspace{-4mm}
\end{table}

\subsection{Deployment Scenarios}\label{Sub a}
We analyse four different RAN configuration scenarios based on the location of the BBP in the RAN architecture. 

\begin{itemize}

  \item \textit{Scenario 1}: In this scenario, BBP is performed at the RU under a D-RAN option, where the full protocol stack is executed locally at the radio site. This represents a largely macrocell configuration with the gNB at the cell site. 
  
  \item \textit{Scenario 2}: This is a typical O-RAN implementation where the BBP is performed at the O-DU. In this configuration, the O-RU and O-DU are co-located or within a short PtP connection of each other. Since baseband functions (RLC/MAC/PHY) reside in the O-DU, the functional separation with respect to the CU follows the standard Split 7.2 option (at RU-DU).
  
  \item \textit{Scenario 3}: Here, BBP is centralized at the CU hosted in an edge/regional cloud, with the RU acting as a lightweight RF front end and the remaining protocol stack executed at the CU. This Split 8-like edge deployment represents an emergent RAN architecture that pools processing closer to the RAN than a distant DC, retaining centralization gains while reducing transport distance and helping satisfy latency constraints.
  
  \item \textit{Scenario 4}: In this Split 8 configuration, BBP is fully centralized in the regional DC. The RU remains a lightweight RF front end, while the PHY and all higher protocol layers are executed centrally. This scenario is a C-RAN architecture, with the DU and CU (gNB) located in the DC and all other network nodes forwarding the radio signals through the network.

\end{itemize}

\begin{figure*}[!t] 
    \centering
    \includegraphics[width=7.2in]{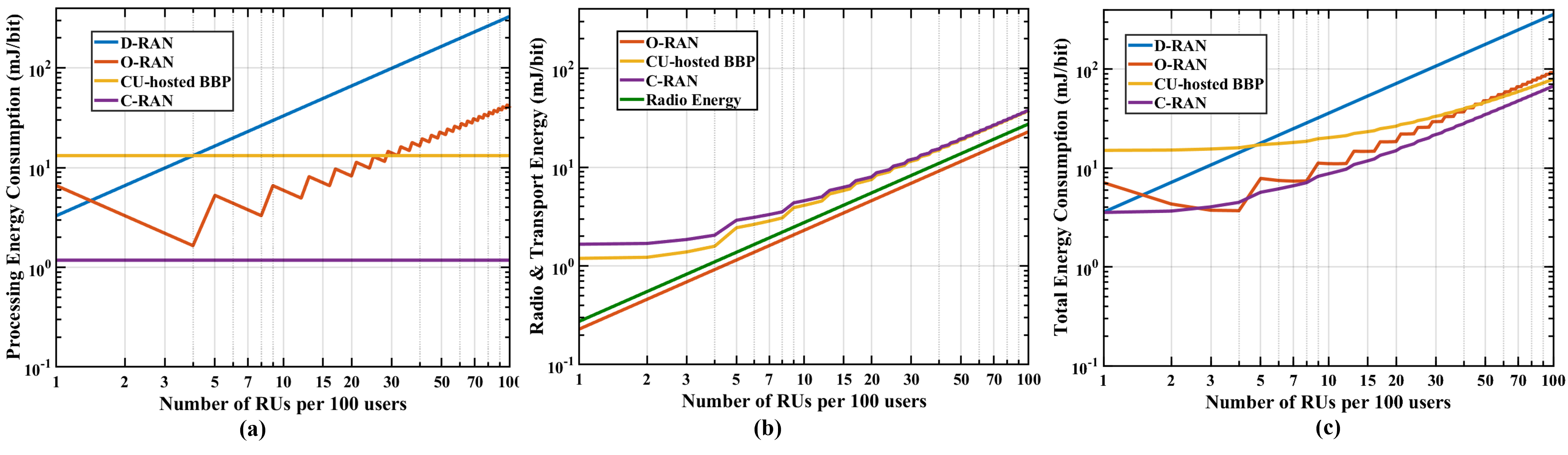}
    \vspace{-4mm}
  \caption{Variation in (a) processing, (b) radio $\&$ transport, and (c) total energy consumption with increase in network densification for different deployment scenarios. In (b), the transport energy for D-RAN (BBP hosted at RU) is not plotted as it is fairly constant at a low value ($\sim$ 40 nJ/bit).} 
    \label{fig:energy}
    \vspace{-4mm}
\end{figure*}

\subsection{Network Densification}\label{Sub b}
 To study network densification, we consider 100 users or CPEs supported by a RAN and increase the number of serving RUs from 1 to 100. We provision one additional DU for every four RUs. This fan-out is commonly supported in open deployments. Larger fan-outs might be supported but imply greater processing oversubscription. Moreover, we assume there are two CUs connected to a single DC in the network core. As shown in Fig. \ref{fig:energy}(a), the processing energy in D-RAN (Scenario-1) rises monotonically with the number of RUs because BBP is performed independently at each RU. In O-RAN (Scenario-2), a newly introduced DU initially serves few RUs, giving relatively high processing energy per bit. As more RUs share the DU, resource utilization improves, and the processing energy decreases. Provisioning another DU after every four RUs produces the sawtooth trend, whose step frequency depends on the number of RUs supported per DU. However, in case of CU-hosted BBP (Scenario-3) and C-RAN (Scenario-4), the processing energy remains approximately constant at $\sim$14 mJ/bit and $\sim$1.2 mJ/bit, respectively, because the number of CUs is fixed at two and BBP in C-RAN is performed at the single regional DC. Thus, D-RAN and C-RAN have the highest and lowest consumption of processing energy, respectively, because C-RAN benefits from multiplexing gain due to the sharing of resources at the core.

In Fig. \ref{fig:energy}(b), we depict the scenario-independent radio energy and transport energy for Scenarios 2–4. In D-RAN, eCPRI data is processed at the RUs, so transmission energy is nearly constant with RU densification at a low value of $\sim$40 nJ/bit. In all other scenarios, the transport energy consumption is higher and scales with densification because more eCPRI traffic from the growing number of RUs must reach a BBP node located further into the network. In case of CU-hosted BBP and C-RAN, the eCPRI data hops through DUs, thereby producing step-like curves for the same reason as in Fig. \ref{fig:energy}(a). The radio energy also increases with densification because additional radio equipment must be provisioned as the number of RUs increases. Finally, the total energy consumption per user shown in Fig. \ref{fig:energy}(c) is essentially the sum of the components depicted in Figs. \ref{fig:energy}(a) and (b). At very low densification (\(N_{\mathrm{RU}}\)$<$5), the energy consumption in the case of CU-hosted BBP is higher because the fixed processing consumption of two CUs is shared by few RUs. With increasing densification, D-RAN rapidly becomes the least energy-efficient because its processing energy grows strongly with the number of RUs. O-RAN initially consumes less total energy than Scenario-3 (CU-hosted BBP), but overtakes it at \(N_{\mathrm{RU}}\) $\approx$ 30 as additional DUs are provisioned. C-RAN provides the lowest total energy consumption over most of the densification range because its processing energy remains very low despite its higher transport energy. Thus, with one RU per user, D-RAN consumes $\sim$74–81\% more energy than the other three cases, with C-RAN providing the largest saving. Overall, increasing centralization shifts the dominant energy component from processing toward transport. These counterbalancing trends explain why the total energy curve is most sensitive to the BBP location, and they reinforce the conclusion that intelligent placement, not merely densification, is what determines whether centralization actually reduces the network’s energy footprint.

\section{Conclusion}
In this work, we explore the energy consumption of emergent RAN architectures. In particular, we update past results using today's equipment and compare access technology energy efficiency. We also look at the historical efficiency of processor-dominant network equipment (edge routers) and find an efficiency improvement rate that is in line with predictions (and nevertheless slower than traffic growth rates). To consider today's open and virtualized RANs, we need to include the full network energy from the access point to the data center. By including this end-to-end aspect, we show that the relative contributions of processing and transport energy depend strongly on the location of BBP and the level of network densification. Depending on the deployment strategy, greater sharing of the processing load is shown to yield significant energy efficiency improvements.  These findings highlight that the placement of baseband processing within the network plays a crucial role in determining overall energy efficiency. While centralized architectures like C-RAN consume higher transmission energy due to longer data transport, their superior computational efficiency can offset this cost. Ultimately, designing future communication networks will require a careful balance between processing distribution and transmission distance to achieve sustainable energy performance.

\section*{Acknowledgments}
\small{This work was supported in part by Research Ireland grants 18/CRT/6222 and 13/RC/2077\_P2; and by the EU projects 6G-XCEL (101139194), ECO-eNET (101139133) and ENGCoN (101155602).}

\bibliographystyle{ieeetr}
\bibliography{references}

\end{document}